\documentstyle[11pt,newpasp,twoside]{article}
\input epsf.sty
\def\edcomment#1{\iffalse\marginpar{\raggedright\sl#1\/}\else\relax\fi}
\reversemarginpar

\begin{document}
\title{Detection of new emission components in PSR~B0329+54}
 \author{R. T. Gangadhara}
\affil{Indian Institute of Astrophysics, Bangalore - 560034, India}
\author{Y. Gupta}
\affil{National Centre for Radio Astrophysics, TIFR,  
   Pune - 411007, India }
\author{D. R. Lorimer}
\affil{Arecibo Observatory, HC3 Box 53995, Arecibo, PR 00612, USA}
\begin{abstract}
To study the structure of emission beam, we have analysed the
single pulse data of PSR~B0329+54 at 325 and 606~MHz.  In order to
unambiguously detect the weak emission components, we have developed a 
new data analysis technique, which we
term ``window--thresholding''. By applying this technique to the data,
we have detected three new emission components, and also
confirmed the presence of a component which was proposed
earlier. Hence our analysis indicates that PSR~B0329+54 has nine
components, which is among the highest of all the known pulsars.  The
symmetric distribution of pulse components about the pulse centre,
indicates that the emission beam is conal.
\end{abstract}
\section{Data Analysis}
For the strong pulsar PSR~B0329+54, the high-quality single pulses are 
easily observable.
We obtained the data at 325~MHz  in March 1999 using the Giant Metrewave
Radio Telescope near Pune, and the data at 606 MHz was obtained in 
August 1996 using the 76-m Lovell Telescope at Jodrell Bank.
For the analysis, we considered about 2500 single 
pulses at each frequency. The time resolutions of data at 325 
and 606 MHz were 0.516 and 0.250 ms, respectively.

To estimate the pulse profiles which clearly show the presence of
weaker components we developed a `window-threshold' technique. In this
technique,
we set a window in longitude and employ an intensity threshold while
considering the single pulses for making average profiles, i.e.~we consider
all those pulses, which have intensity levels above the threshold within the
window.  For threshold the {\em rms} in intensity was 
computed from the off-pulse region.
 In Figs.~1a\&b we have plotted the average pulse profiles which 
were obtained by applying this technique to nine component windows. The average 
profile obtained from all those pulses is plotted in Figs.~1c\&d, which 
clearly shows the presence of 9~emission components (I, II, III, IV, V, VI, VII,
VIII, IX) in PSR~B0329+54. The components on trailing side of the profile are 
closely spaced compared to the leading side.
\begin{figure}
\epsfysize8.80 truecm\centerline{{\epsffile[ 0 0 540 540]{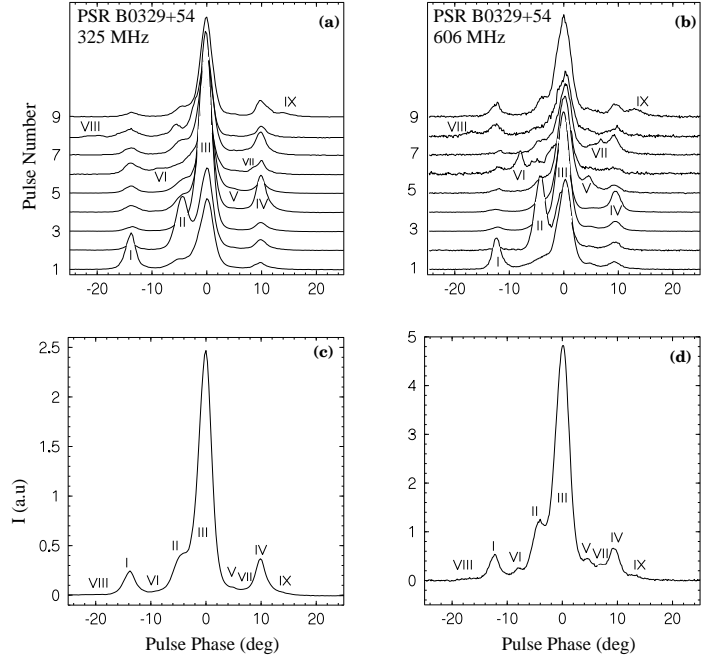}}}
\caption{(a)\&(b) are the average profiles obtained by using the 
window-threshold 
technique for each component at 325 and 606 MHz, while (c)\&(d) are their
average profiles with arbitrary units (a.u).}
\end{figure}
\section{Conclusion}
  We have developed a technique based on windowing and thresholding,
to detect the weak emission components in pulsar profiles. By applying it
to the single pulse data of PSR~B0329+54 we have detected three new emission
components (VII, VIII and IX) of this pulsar, and also confirmed the
presence of a component (VI) proposed by Kuzmin \& Izvekova (1996). 
The near-symmetric distribution of components around the
core/centre of the profile, favours the idea that the pulsar emission beam
is annular or conal (Oster \& Sieber 1977; Rankin 1983).
\acknowledgments
We thank A. G. Lyne for providing the Jodrell Bank data, and
J. M. Rankin for useful discussions.

\end{document}